\documentstyle[11pt,paspconf,epsf]{article}
\begin{document}
\title{Hard X-ray-to-radio energy distributions in starburst galaxies}

\author{M. Cervi\~no,  J.M. Mas-Hesse}
\affil{LAEFF-INTA, P.O. Box 50727, E-28080 Madrid, Spain.}

\begin{abstract}
We present in this contribution the predictions on the multiwavelength
spectral energy distribution of our evolutionary population synthesis
models including single and binary stellar systems. The high energy
computations include the emission associated to X-ray binaries and
supernovae remnants, as well as the mechanical energy released into the
interstellar medium, which can be partially reprocessed into thermal
X-rays. With these components we compute the spectral energy distribution
of starburst galaxies from X-ray to radio ranges, and analyze finally the
effects of the high energy emission on the H and He ionizing continuum. 
\end{abstract}


\section{Predictions on spectral energy distribution (SED)}

Our models including the evolution of single and binary stars have been
discussed in an accompanying contribution in this volume (Mas-Hesse and
Cervi\~no, ``Evolutionary population synthesis: the effect of binary
systems''). Here we present the predictions corresponding to the
multiwavelength SED and the He and H emission lines. 
The main contribution of this work is the inclusion of the X-ray domain in
the multiwavelength energy distribution of evolutionary synthesis models.
The basic components of the X-ray emission included in the code are:

\begin{itemize} 
\item {\it Mechanical energy} released by SN and stellar winds in the ISM,
  which is partially reprocessed to thermal soft X-rays. 
\item {\it Supernovae remnants}, which have been assumed to generate
 two components with a composed  Raymond-Smith spectrum.
\item {\it High mass X-ray binaries}, with circular orbits (i.e. giant or
  supergiant stars) and stellar wind accretion onto a compact object,
  producing hard black body emission.
\end{itemize}

We show in Fig.~1 the predicted SED for 2 ages, assuming an instantaneous
burst. A very young cluster (0.25~Myr) shows thermal radio emission, some
thermal soft X-rays contribution from the hot, diffuse interstellar gas and
almost no emission at hard X-rays. At 16~Myr, however, the radio emission
has become non-thermal and is dominated by supernova remnants. Supernova
remnants also contribute to the soft X-ray emission, and a strong hard
X-ray component associated to High-Mass-X-ray binaries appears. The Far
Infrared emission in the IRAS band is due to dust reprocessing thermally
the absorbed UV photons.

\begin{figure}
\begin{center}
\plotfiddle{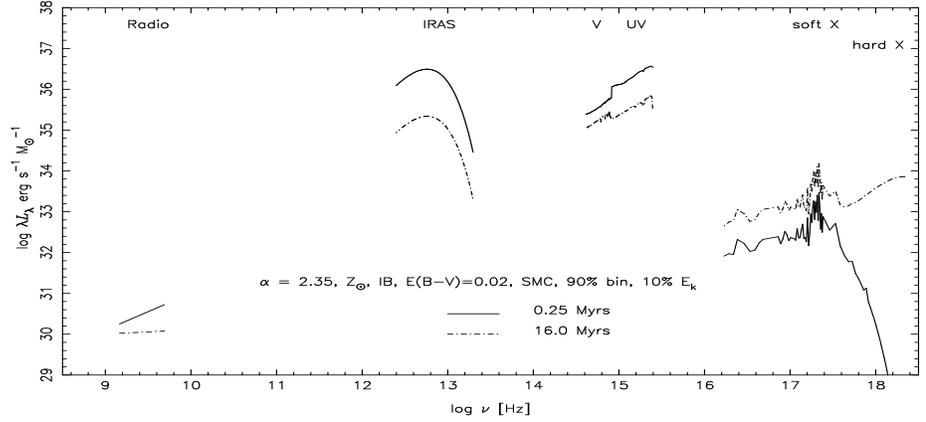}{4.0cm}{0.}{80.}{58.}{-210.}{-30.}
\end{center}
\caption{Multivawelength spectral energy distributions at 2 ages.}
\end{figure}

\section{The effects on the emission lines}

The effect of binary evolution on the H emission lines is shown in
Fig.~2a. It can be seen that the ionizing flux (and therefore $W(H\beta)$
as well) decreases much slower than in a single-star-only cluster, due to
the formation of hot WR and OB stars by mass transfer at ages above 5~Myr.
On the other hand, the additional ionizing photons produced by binary
systems and by the hot interstellar gas increase significantly the
$L(HeII(4686$~\AA$)/L(H\beta)$ ratio (by more than a factor 10 at ages 
larger than 5
Myrs (see Fig. 2b)). Assuming a high efficiency in the reprocessing of
mechanical energy into soft X-rays (top line in Fig. 2b), it is possible to
reproduce the observational values of this ratio, which shows an average
around $L(HeII)/L(H\beta) \approx 0.01$. We conclude that this additional
component of the ionizing flux might be (partially) responsible of the
relatively large $L(HeII)/L(H\beta)$ values observed in several starburst
galaxies.

\begin{figure}
%\begin{center}
\plotfiddle{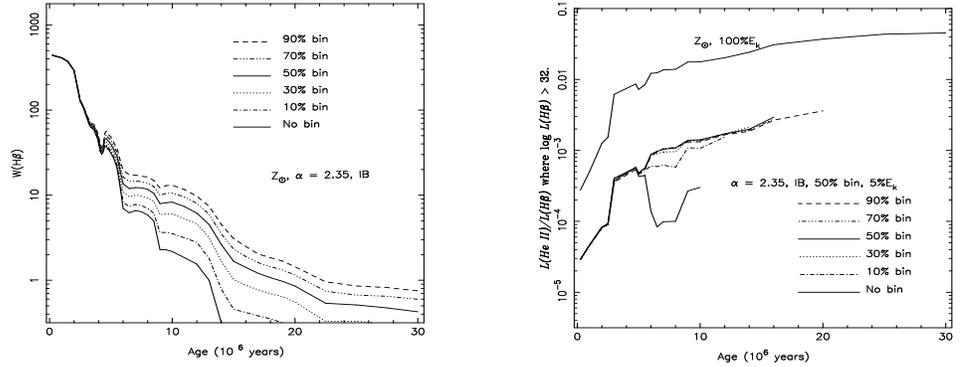}{4.2cm}{0.}{65.}{55.}{-180.}{0.}
%\end{center}
\caption{EW(H$\beta$) and $L(HeII)/L(H\beta)$ as a  function of the
  frequency of binary systems. The top line in (b) has been computed assuming a
  100\% efficiency in the conversion of mechanical energy into soft X-rays,
  which contribute to the ionization of HeII.}
\end{figure}

\end{document}